\documentclass[12pt]{iopart}
 \pdfoutput=1 
\usepackage{graphicx}
\usepackage{amssymb}
\usepackage{color}
\usepackage{cite}   
\DeclareGraphicsExtensions{.pdf,.png,.jpg} 

\begin{document}
\newcommand{\wq}[1]{\textcolor{blue}{#1}}

\title{Single-file diffusion on self-similar substrates}

\author{G.~P.~Su\'arez, H.~O.~M\'artin, J.~L.~Iguain}
\address{Instituto de Investigaciones F\'{\i}sicas de Mar del Plata (IFIMAR) 
and\\
Departamento de F\'{\i}sica FCEyN,
Universidad Nacional de Mar del Plata,\\
De\'an Funes 3350, 7600 Mar del Plata, Argentina}


\begin{abstract}
We study the single file diffusion problem on a one-dimensional lattice with a self-similar distribution of hopping rates. We find that the time dependence of the mean-square displacement of both a tagged particle and the center of mass of the system present anomalous power laws modulated by logarithmic periodic oscillations. The anomalous exponent of a tagged particle is one half of the exponent of the center of mass, and always smaller than $1/4$. Using heuristic arguments, the exponents and the period of oscillations are analytically obtained and confirmed by Monte Carlo simulations.
\end{abstract}
\maketitle

\section{Introduction}

  The origin of the anomalous diffusion on fractal substrates has been well understood several decades ago (see, \textit{e.g.}, Ref.\cite{havlin1987,bouchaud1990,rammal1983,alexander1982,ben2000diffusion}). However, in the last years, it has been repeatedly reported that, on certain kind of self-similar objects, the anomalous diffusion is modulated by logarithmic-periodic oscillations (see, \textit{e.g.}, Ref.\cite{Grabner1997, acedo2000, bab2008EPL, bab2008JCP, maltz2008, weber2010, lorena2010,  bernhard2004}).  In all these cases, the time behavior of a single random walk (RW) or a set of noninteracting RWs were studied. The origin of the anomalous diffusion (with or without oscillations) can be found in the self-similar character of the substrate where the particles move.
  
  It is also well known that the anomalous diffusion can also appear in a simple one-dimensional lattice if hard-core interaction between particles is considered, i.e., two particles cannot occupy the same lattice site, and a particle cannot cross over another one (the so called single-file diffusion). More specifically, the mean-square displacement (MSD) of a tagged particle behaves as $\Delta^2x_{HC}\sim t^{2\nu_{HC}}$ (with $2\nu_{HC}=1/2$) after a crossover time, which depends on the concentration of particles (see, \textit{e.g.}, Ref. \cite{harris,richards1977,beijeren1983,lizana2009,manzi2012,centres2010}). Note that, despite hard-core interactions, the center of mass of the system presents a normal behavior. This model, and similar ones, are useful to describe several microscopic processes in physics, chemistry and biology (see, \textit{e.g.}, Ref. \cite{Wei2000,zeolites,biological}). 

  The aim of the present work is to study the single-file diffusion problem on the one-dimensional lattice with a self-similar distribution of hopping rates introduced in Ref. \cite{Padilla2009}. We find that, for a tagged particle, the MSD presents a global anomalous behavior $\Delta^2x_{HC}\sim t^{2\nu_{HC}}$ (with $2\nu_{HC}<1/2$) modulated by log-periodic oscillations. The center of mass also follows an anomalous behavior modulated by oscillations.

The paper is organized as follows. In Sec.~\ref{sec:substrate} we describe the substrate where the diffusion takes place. In Sec.~\ref{sec:non-interacting} we review the analytical solution found for non-interacting particles diffusing in this substrate \cite{Padilla2009}. In Sec.~\ref{sec:hard-core} we derive analytical solutions for the case of many particles interacting through a hard-core potential; we analyzed both the time evolution of the center of mass (Sec.~\ref{sec:hard-core-cm}) and that of a tagged particle (Sec.~\ref{sec:hard-core-tagged}).
In Sec.~\ref{sec:results}, we show the results of numerical simulations and compare them with our theoretical predictions; in the oscillatory regime (Sec.~\ref{sec:results-osc}), and in the long time regime (Sec.~\ref{sec:results-long-time}). Some final remarks are exposed in Sec.~\ref{sec:conclusion}.  

\section{The substrate} \label{sec:substrate}

\begin{figure}[ht]
  \begin{center}
    \includegraphics[scale=.35,clip]{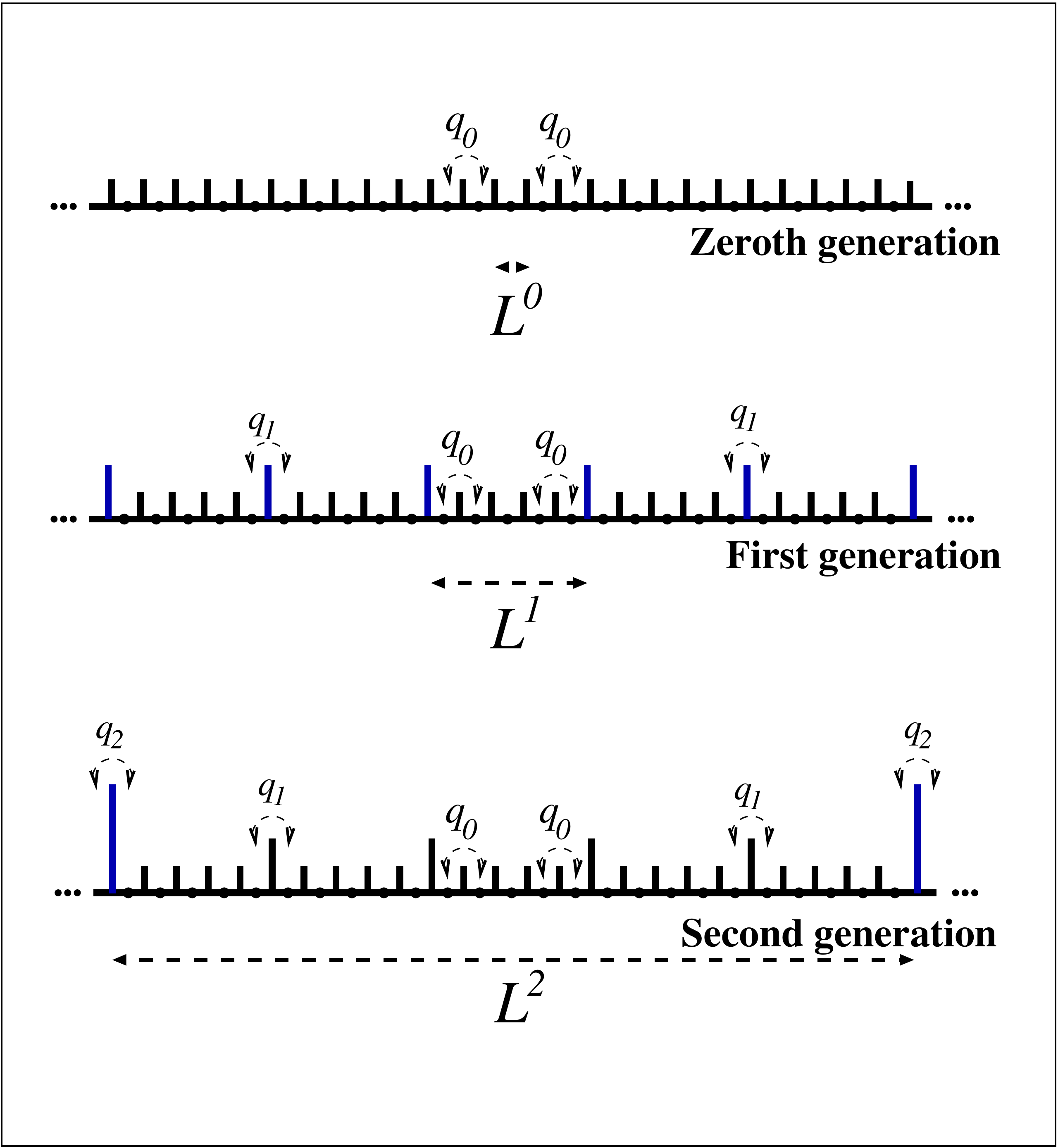}
    \caption{Construction of a lattice with a self-similar distribution of hopping rates ($L=5$). The lattice sites $i\:(i\in \mathbb{Z})$ are located between consecutive barriers. The site $i=0$ is located in the center of the scheme. Top: The zeroth generation ($n=0$) with all the hopping rates equal to $q_0$. Center: The first generation ($n=1$), where the hopping rates $q_1$ appear. The corresponding barriers are separated by a distance $L$. Bottom: The second generation ($n=2$) shows the emergence of the hopping rates $q_2$ separated by a distance $L^2$. In the limit \mbox{$n\rightarrow\infty$} a self-similar distribution of hopping rates is obtained. For more details, see the text.}
  \label{substra}
  \end{center}

\end{figure}

  We use the self-similar one-dimensional model introduced in Ref.~\cite{Padilla2009}, which depends on two parameters $L$ (integer greater than $2$) and $\lambda$ (real positive). See Fig.~\ref{substra} and this reference for further details.
  The particles only jump between nearest-neighbor sites of the lattice. The substrate where the particles diffuse is built in stages and the result of every stage is called a {\it generation}. In the zeroth-generation substrate, all the hopping rates are identical ($q_0$). In the first generation, the hopping rates that correspond to every lattice site $j=pL-(L+1)/2$, with $p$ integer, are set to $q_1$ ($<q_0$), while the other hopping rates remain as in the generation zero.
  This process is iterated indefinitely and, in general, the generation $n$ is obtained from the generation $n-1$ after replacing by $q_n$ ($<q_{n-1}$) the values of the hopping rates which correspond to every lattice site \mbox{$j=pL^n-(L^n+1)/2$}, with $p$ integer. In the limit of an infinite number of iterations ($n \rightarrow \infty$) we get the model, with a distribution of hopping rates $q_i$, ($i=1,2,...$) defined by
\begin{equation}
  \begin{centering}
    \frac{q_{0}}{q_{i}}= \frac {q_{0}}{q_{i-1}} + (1+\lambda )^{i-1}
    \lambda L^{i},\;\;\;{\rm for}\; i=1,2,3...\;\;\; .
  \end{centering} \label{relation}
\end{equation}

\section{Non-interacting particles} \label{sec:non-interacting}

In order to introduce the procedure used to obtain expressions for the period of
the oscillations and the exponent of the MSD, we begin by analyzing the case of non-interacting particles.

As shown in Ref. \cite{Padilla2009}, on the substrate obtained after an infinite number of iterations, the diffusion of a single particle is perfectly self-similar. Then, if we consider a set non-interacting particles performing independent RWs, for length scales between $\sim\!\!L^{n}$ and $\sim\!\!L^{n+1}$ ($n=0,1,2,...$), everything occurs as in the periodic lattice of the generation $n$, and the MSD of any tagged particle of the sample $\Delta^2x_{\mbox{\tiny NI}}=\langle [x_{\mbox{\tiny NI}}(t) - x_{\mbox{\tiny NI}}(0)]^2 \rangle$ (where $\langle ... \rangle$ denotes average over RW realizations) as a function of time $t$ satisfies 
\begin{equation} 
  \Delta^2x_{\mbox{\tiny NI}}=2 D_{\mbox{\tiny NI}}^{(n)}t,
  \label{msd_perio_ni}
\end{equation}

\noindent where 
\begin{equation}
  D_{\mbox{\tiny NI}}^{(n)}=q_0/(1+\lambda)^n
  \label{dif_coe_ni}
\end{equation}

\noindent is the single particle diffusion constant of the \mbox{$n^{\rm th}$} generation substrate. More generally, because of the infinite set of diffusion constants, \mbox{$\{ D_{\mbox{\tiny NI}}^{(n)}; n=0,1,2,...\}$},  for a time $t>1/q_0,$ the behavior of $\Delta^2x_{\mbox{\tiny NI}}(t)$ is well described by
\begin{equation}
  \begin{centering}
    \Delta^2x_{\mbox{\tiny NI}}(t) =  t^{2 \nu} f(t),
  \end{centering}
  \label{function}
\end{equation}

\noindent where $f(t)$ is a log-periodic function, which satisfies $f(t \tau)=f(t)$. The values of the single particle RW exponent $\nu$ and the logarithmic period $\tau$ are, respectively~\cite{Padilla2009},
\begin{equation}
  \nu = \frac{1}{2+\frac{\log(1+\lambda)}{\log L}}\;\;\;{\rm ,}
  \label{exponent}
\end{equation}

\noindent and
\begin{equation}
  \tau=(1+\lambda)L^2\;\;{\rm .}
  \label{tau}
\end{equation}

\section{Hard-core interactions} \label{sec:hard-core}

If the $N$ particles, which diffuse on the self-similar substrate described, above are not independent but interact with each other, the problem is considerably more complex. In what follows we study the case of hard-core interactions. That is, two particles cannot occupy the same lattice site and a particle cannot cross over another particle. We address first the behavior of the center of mass and then that of a tagged particle.

\subsection{Center of mass} \label{sec:hard-core-cm}

In spite of the interactions, the center of mass of the system $\overline{x}(t)=(1/N)\sum_{i=1}^N x_i(t)$ mirrors the behavior of an independent particle. For 
any one-dimensional periodical substrate of period $\ell$, the center-of-mass MSD $\Delta^2 \overline{x}(t)=\langle [\overline{x}_{\mbox{\tiny NI}}(t) - \overline{x}_{\mbox{\tiny NI}}(0)]^2 \rangle$ is directly proportional to time $t$ and
and inversely proportional to $N$, i.e., 
\begin{equation}
N  \Delta^2 \overline{x}(t)= 2 \overline{D} t;
  \label{cm_perio}  
\end{equation}
a valid relation for times such that $\sqrt{\Delta^2 \overline{x}(t)} > \gamma \ell$. That is, normal diffusion should be observed if time is long enough for the RW to be influenced by the structure periodicity. The constant $\gamma$ is, roughly speaking, the fraction of  $\ell$ the center of mass has moved when start realizing the substrate is periodic. Both $\gamma$ and the coefficient $\overline{D}$ in Eq.~(\ref{cm_perio}) depend on the concentration, $c$, of particles and on the manner they interact. 

\begin{figure}[ht]
  \begin{center}
    \includegraphics[scale=1.1,clip]{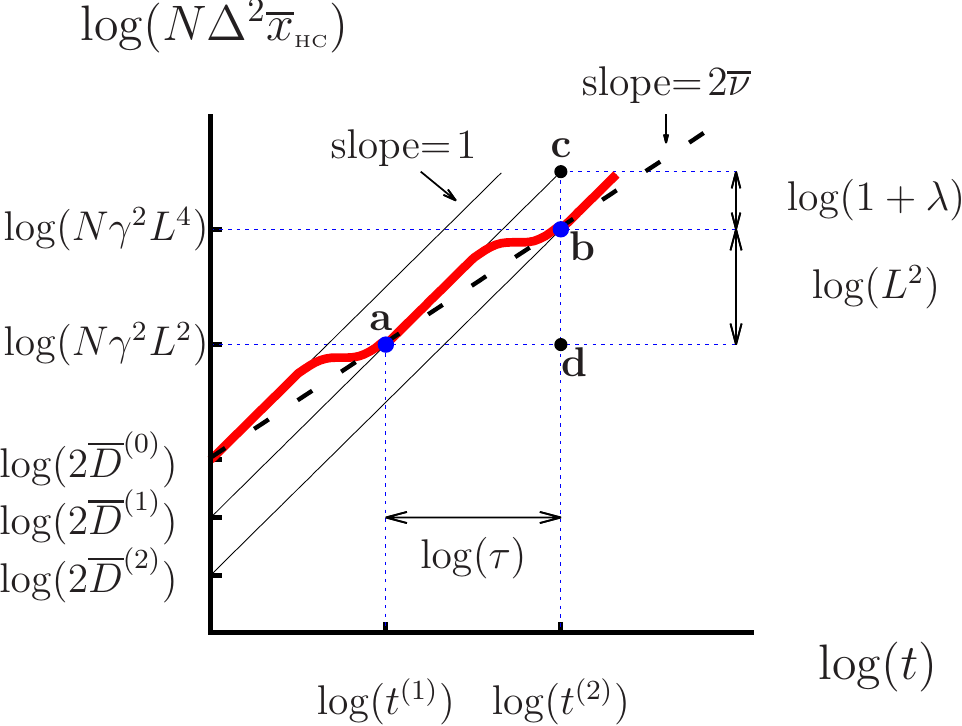}
  \end{center}
  \caption{(Color online) Schematic of the center-of-mass MSD as a function of time, shown by the thick red curve. The length of the segment {\bf bc} is $\log(2\overline{D}^{(1)})-\log(2\overline{D}^{(2)})=\log(1+\lambda)$, because of Eq.~(\ref{coef_cm_hd}). From the slopes ($=1$) of the full straight lines (representing the normal diffusion behaviors, $N\Delta^2\overline{x}_{\mbox{\tiny{HC}}} = 2\overline{D}^{(n)}t$), one gets that the segments {\bf ad} and {\bf cd} have the same length or, equivalently, that $\log\tau=\log L^2+\log(1+\lambda)$. The dashed straight line represents the global power law $\Delta^2\overline{x}_{\mbox{\tiny{HC}}} \sim t^{2\overline{\nu}}$, with $2\overline{\nu}= \log L^2/\log\tau$. More details in the text.}
  \label{esque_cm}
\end{figure}

When the interactions are of hard-core type, it is found that $\overline{D}=D_{\mbox{\tiny NI}} (1-c)$ \cite{Terranova2005} (let us note that the system is equivalent to a chain of $N$ particles), where $D_{\mbox{\tiny NI}}$ is the diffusion coefficient of non-interacting particles on the same substrate. In particular, on the  \mbox{$n^{\rm th}$} generation substrate in Fig.~\ref{substra}, 
\begin{equation}
  \overline{D}^{(n)}=D^{(n)}_{\mbox{\tiny NI}} (1-c)=\frac{(1-c) q_0}{(1+\lambda)^n},
  \label{coef_cm_hd}
\end{equation}

and the MSD of the center of mass will satisfy
\begin{equation}
  N \Delta^2 \overline{x}_{\mbox{\tiny HC}}(t)= 2 \overline{D}^{(n)} t,
  \label{msd_cm_perio}
\end{equation}

\noindent for $t$ long enough, i.e., such that $\Delta^2 \overline{x}_{\mbox{\tiny HC}}(t) > \gamma^2 L^{2n}$ ($L^n$ is the period of this substrate).

For hard-core interacting particles, diffusing on the full self-similar substrate, the center-of-mass MSD will behave as in the $n^{\rm th}$ generation substrate for a time longer than $t^{(n)}$, given by $\Delta^2 \overline{x}_{\mbox{\tiny HC}}(t^{(n)}) \sim \gamma^2 L^{2n}$ but shorter than $t^{(n+1)}$, given by $\Delta^2 \overline{x}_{\mbox{\tiny HC}}(t^{(n+1)}) \sim \gamma^2 L^{2(n+1)}$. For times between $t^{(n+1)}$ and $t^{(n+2)}$, where $\Delta^2 \overline{x}_{\mbox{\tiny HC}}(t^{(n+2)}) \sim \gamma^2 L^{2(n+2)}$, it should happen that $N\Delta^2 \overline{x}_{\mbox{\tiny HC}}(t)= 2 \overline{D}^{(n+1)} t$, and so on. Thus, we expect that the MSD of the center of mass behaves as sketched in Fig.~\ref{esque_cm}, where the thick red curve represents the function $\Delta^2 \overline{x}_{\mbox{\tiny HC}}(t)$. In the same figure, the inclined solid lines represent the normal diffusion of the center of mass in every periodic substrate of the construction in Fig.~\ref{substra} (with a coefficient $\overline{D}^{(n)}$ for the generation $n$), and were drawn to guide the eyes.

According to Eq.~(\ref{coef_cm_hd}), the ratio between consecutive coefficients results
\begin{equation}
  \frac{\overline{D}^{(n)}}{\overline{D}^{(n+1)}}=1+\lambda,
\end{equation}

\noindent which implies, as is evident after simple geometrical analysis of Fig.~\ref{esque_cm}, that both the random walk exponent and the period of the oscillations are the same as for non-interacting particles, i.e.,

\begin{equation}
  \overline{\nu} = \frac{1}{2+\frac{\log(1+\lambda)}{\log L}}\:,
  \label{exponent_cm}
\end{equation}

\noindent and
\begin{equation}
  \tau=(1+\lambda)L^2\:.
  \label{tau_cm}
\end{equation}

\subsection{Tagged particle} \label{sec:hard-core-tagged}

When $N$ hard-core interacting particles diffuse in a one-dimensional medium, single-file diffusion occurs, and for a periodic substrate, after a transient time, the MSD of a tagged particle of the sample satisfies~\cite{beijeren1983}
\begin{equation}
  \Delta^2 x_{\mbox{\tiny HC}}(t)=\frac{2(1-c)a}{c\sqrt{\pi}}\sqrt{D_{\mbox{\tiny NI}}t},
  \label{tag_perio}
\end{equation}
\noindent where $a$ is the lattice spacing.

  In an infinite-size system, this anomalous diffusive behavior with a RW exponent $\nu=1/2$ occurs for ever but, in a real system, normal diffusion of the tagged particle is recovered after a crossover time, which grows with system linear size and decreases with  particle concentration~\cite{centres2010,beijeren1983}.

For the substrate we are interested in, i.e., the self-similar lattice in Fig.~\ref{substra}, the task of capturing the behavior of one tagged particle looks harder. However, the analysis becomes straightforward if we assume that for a time $t$ between $t^{(n)}$ and $t^{(n+1)}$, the whole system behaves as in the $n^{\rm th}$-generation periodic substrate, or, in other words, that every characteristic time $t^{(n)}$ ($n=1,2,...$) plays the role of a crossover between two dynamical regimes; each one identical in average to that observed on some of the periodical substrates in Fig.~\ref{substra}.

\begin{figure}[ht]
  \begin{center}
    \includegraphics[scale=1.1,clip]{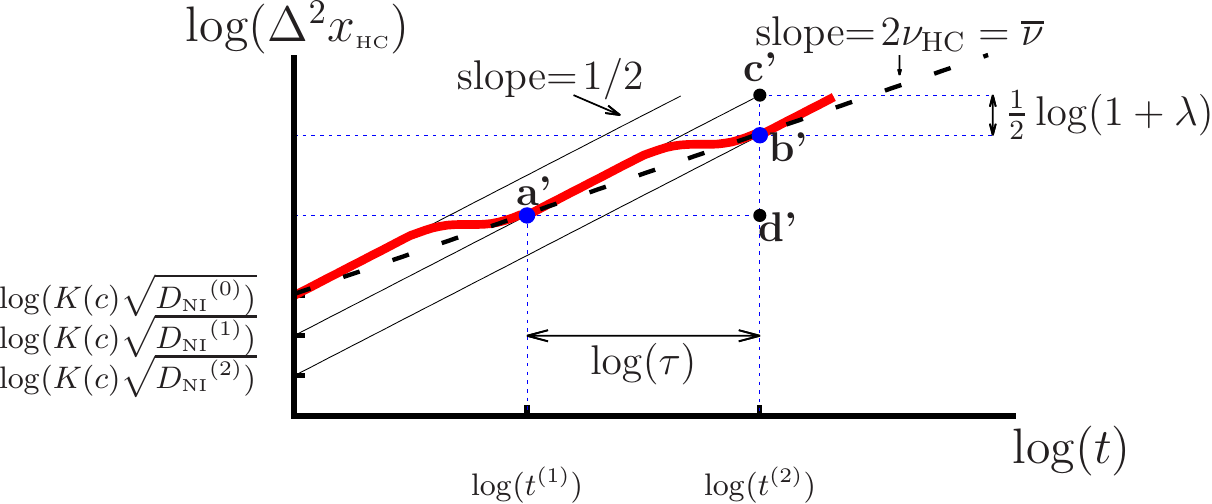}
  \end{center}
  \caption{(Color online) Schematic of the MSD of a tagged particle as a function of time, shown by the thick red curve. The length of the segment {\bf b'c'} is $\log\left(K(c)\sqrt{D_{\mbox{\tiny NI}}^{(1)}}\right)-\log\left(K(c)\sqrt{D_{\mbox{\tiny NI}}^{(2)}}\right)=\log\left(1+\lambda\right)/2$, because of Eq.~(\ref{coef_cm_hd}). From the slopes ($=1/2$) of the full straight lines (representing the tagged particle behaviors, $\Delta^2x_{\mbox{\tiny HC}}(t) = K(c) \sqrt{D_{\mbox{\tiny NI}}^{(n)}t}$, with $K(c)=\frac{2(1-c)a}{c\sqrt{\pi}}$), we obtain  that the length of the segment {\bf c'd'} is half of that {\bf a'd'}. The dashed straight line represents the overall trend $\Delta^2x_{\mbox{\tiny HC}} \sim t^{2\overline{\nu}_{\mbox{\tiny HC}}}$, with $2\nu_{\mbox{\tiny HC}}= \overline{\nu}$.}
  \label{esque_hc}
\end{figure}

With this assumption, and according to Eq.~(\ref{tag_perio}), the MSD of a tagged particle  will be
\begin{equation}
  \Delta^2 x_{\mbox{\tiny HC}}(t)=\frac{2(1-c)a}{c\sqrt{\pi}}\sqrt{D_{NI}^{(n)}t},\;\;\;\;\; \mbox{for}\;\; t\;\; \mbox{between}\;\; t^{(n)}\;\; \mbox{and}\;\; t^{(n+1)},
  \label{msd_hc}
\end{equation}

\noindent and more generally, the form of $\Delta^2 x_{\mbox{\tiny HC}}(t)$ will be as sketched in Fig.~\ref{esque_hc}; a qualitative plot of a global power-law trend modulated by a log-periodic function, i.e., 
\begin{equation}
  \Delta^2 x_{\mbox{\tiny HC}}(t)=t^{2\nu_{\mbox{\tiny HC}}}g(t), \;\;\;\;\;\; \mbox{with}\;\;\; g(t\tau)=g(t). 
  \label{log_per_tag}
\end{equation}

In Fig.~\ref{esque_hc}, the inclined solid line with slope of $1/2$ correspond to the MSD of a tagged particle in the different generations, which approximate  the function $\Delta^2 x_{\mbox{\tiny HC}}(t)$ on the perfect self-similar substrate when $t$ lies in the appropriate time window. In Eq.~(\ref{log_per_tag}), we have made explicit that $g(t)$ is a log-periodic function with the same logarithmic period $\tau$ of $f(t)$ (Eqs.~(\ref{function}) and (\ref{tau})), as is evident from the fact that for both the center of mass and the tagged particle $\tau=t^{(n+1)}/t^{(n)}$. The dashed straight line stands for the overall general trend, i.e., $\sim t^{2\nu_{\mbox{\tiny HC}}}$. As for the center of mass case, the exponent $\nu_{\mbox{\tiny HC}}$ as a function of $L$ and $\lambda$ is dictated by the geometry of this figure:

\begin{equation}
  2\nu_{\mbox{\tiny HC}}=\frac{\frac{1}{2}\log(\tau)-\frac{1}{2}\log(1+\lambda)}{\log(\tau)},
  \label{2nu_tag}
\end{equation}

\noindent or, according to Eq.~(\ref{tau}),
\begin{equation}
  \nu_{\mbox{\tiny HC}}=\frac{\overline{\nu}}{2}.
  \label{nu_tag}
\end{equation}

Thus, we conclude that the MSD of a tagged particle shows a general power-law trend (with an exponent that is half that of the center of mass) modulated by log-periodic oscillations (with the same period as the center of mass).

\section{Numerical results} \label{sec:results}
\subsection{Oscillatory regime} \label{sec:results-osc}
To check the validity of the analytical predictions stated above, we have performed  Monte Carlo (MC) simulations, with $q_0=1/2$ and the distance between nearest-neighbor lattice sites $a=1$. Each lattice site can either be occupied by only one particle or empty. We used a sixth-generation lattice ($L^6$ sites) with periodic boundary conditions. This is a self-similar substrate up to a linear scale of $L^6$. At $t=0$, $N$ particles ($N=cL^6$) are distributed at random on the lattice. They then evolve according to the corresponding hopping rates $q_i, i=0,...,5$. Because of hard-core interactions, if a particle jumps to an occupied site, it returns to the previous position. The simulations were performed up to time $t=10^6$ and the lattices used were large enough to avoid that the MSD of the center of mass reaches the normal diffusion regime given by Eq.~(\ref{msd_cm_perio}) with $n=6$. 

\begin{figure}
  \begin{center}
 \includegraphics[scale=.45,clip]{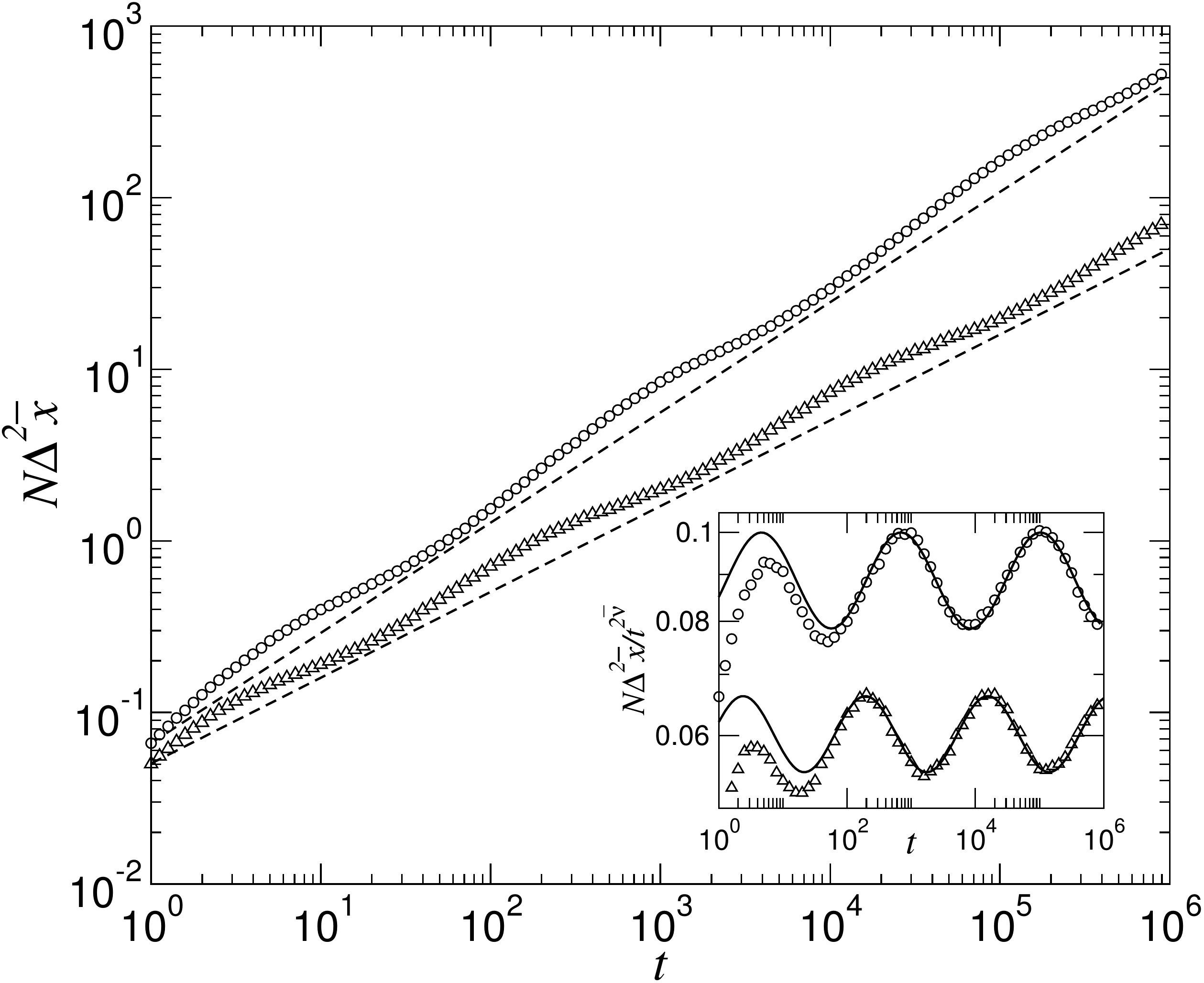}
  \end{center}
  \caption{The MSD of the center of mass as a function of time. Circles: $L=5$ and $\lambda=5$; which give the analytical overall exponent $2\overline{\nu}=0.642$. Triangles: $L=3$, $\lambda=8$; giving $2\overline{\nu}=0.5$. We used $c=0.9$ and an average over $30000$ trajectories in both cases. Straight lines with the theoretically obtained slopes were drawn to guide the eyes. Inset: Is the same data, but with scaled axis. The solid line is a sine function with the theoretical period $\tau$ found. Phase and amplitude are fitted parameters.}
  \label{simu_cm}
\end{figure}

\begin{figure}
  \begin{center}
\includegraphics[scale=.45,clip]{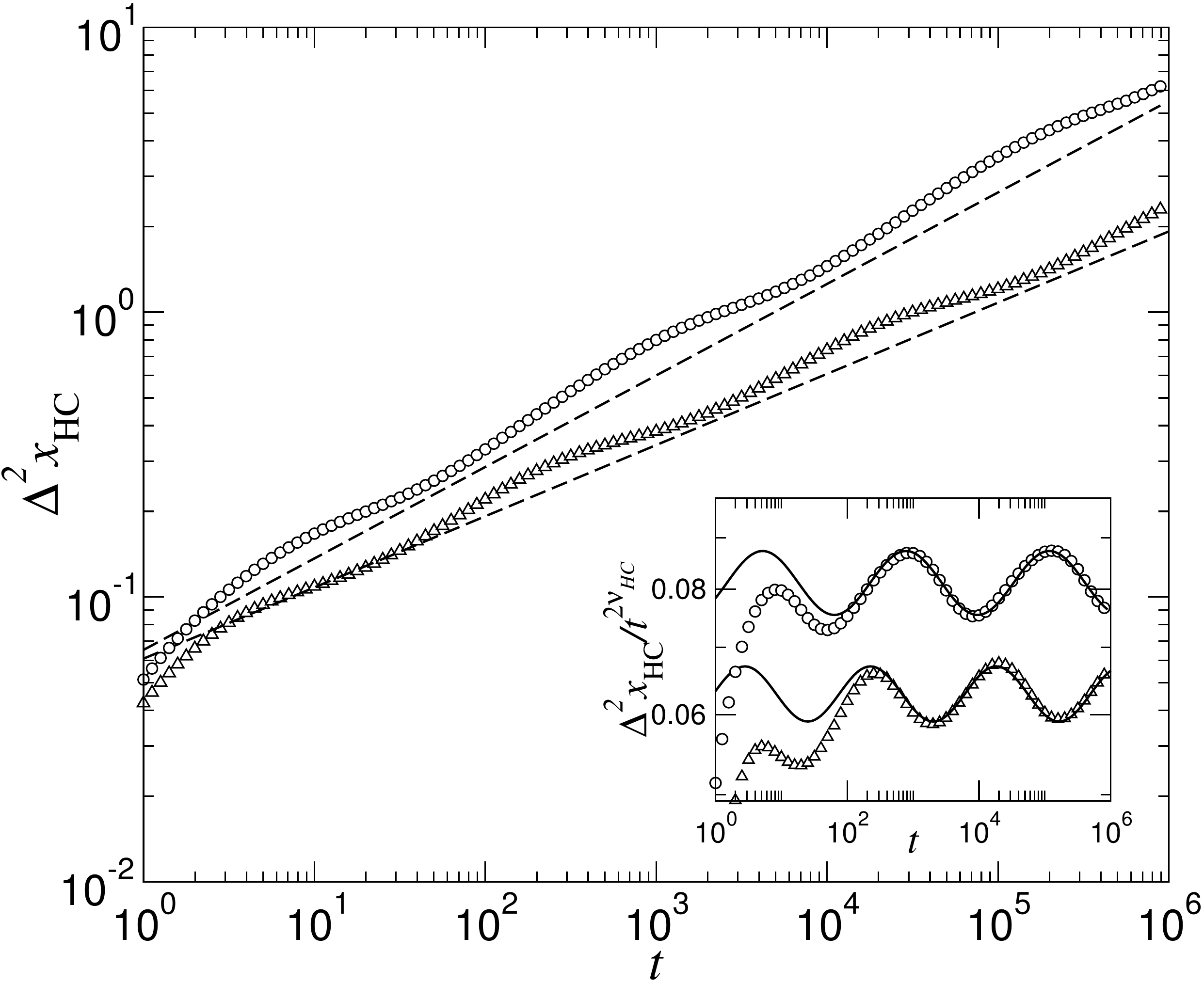}
  \end{center}
  \caption{The MSD of a tagged particle as a function of time. Circles: $L=5$ and $\lambda=5$; which give the analytical overall exponent $2\nu_{HC}=\overline{\nu}=0.321$. Triangles: $L=3$, $\lambda=8$; giving $2\nu_{HC}=\overline{\nu}=0.25$. We used $c=0.9$ and an average over $2\times 10^7$ trajectories in both cases (all the particles of each sample were used as a tagged particle). Inset: Is the same data, but with scaled axis. The solid line is a sine function with the theoretical period $\tau$ given by Eq.~\ref{tau_cm}. Phase and amplitude are fitted parameters.}
  \label{simu_hc}
\end{figure}

The numerical results of the center-of-mass MSD as a function of time is plotted in Fig.~\ref{simu_cm}  for two sets of parameters, $L=5$, $\lambda=5$ and $L=3$, $\lambda=8$. The dashed straight lines have slopes $2\overline{\nu}$ (with the theoretical values from Eq.~(\ref{exponent_cm})) and serve to confirm that $\displaystyle \Delta^2 \overline{x}(t)$ satisfy  modulated power laws with these exponents leading the general trend. The modulations are better observed in the inset, where we have plotted $N \Delta^2\overline{x} / t^{2\overline{\nu}}$ versus $t$, using the same data as in the main plot. Those curves were rigidly displaced in the vertical direction to avoid undesirable superposition of the data points. The curvilinear lines are of the form $\displaystyle A \sin(2\pi \log(t)/\log (\tau) + \alpha)$, with $A$ and $\alpha$ fitted parameters, which corresponds to the first-harmonic approximation of a periodic function, with period $\log(\tau)$. Note the very good agreement between theory and simulations.

The corresponding Monte Carlo results for the tagged particle are plotted in Fig.~\ref{simu_hc}. There, the dashed straight lines represent the theoretical overall behavior of $\Delta^2 x_{HC}(t)$. Their slopes are $2\nu_{HC}=\overline{\nu}$ (Eq.~(\ref{nu_tag})), and they were drawn to guide the eyes.

In the  inset we have plotted $\Delta^2{x}_{HC} / t^{\overline{\nu}}$ as a function of time with the data of the main plot, and sinusoidal curves with logarithmic period $\tau$ Eq.~(\ref{tau}). Note again the very good theory-simulation agreement.

At very short times the effects of hard-core interactions are not completely present. The Monte Carlo results are different from the predicted analytical behavior (See the insets of Fig.~\ref{simu_cm} and Fig.~\ref{simu_hc}). As expected, the numerical results not shown here indicate that the length of this transition regime increases when the concentration $c$ decreases.

\subsection{Long time regime} \label{sec:results-long-time}

\begin{figure}
  \begin{center}
    \includegraphics[width=0.7\textwidth]{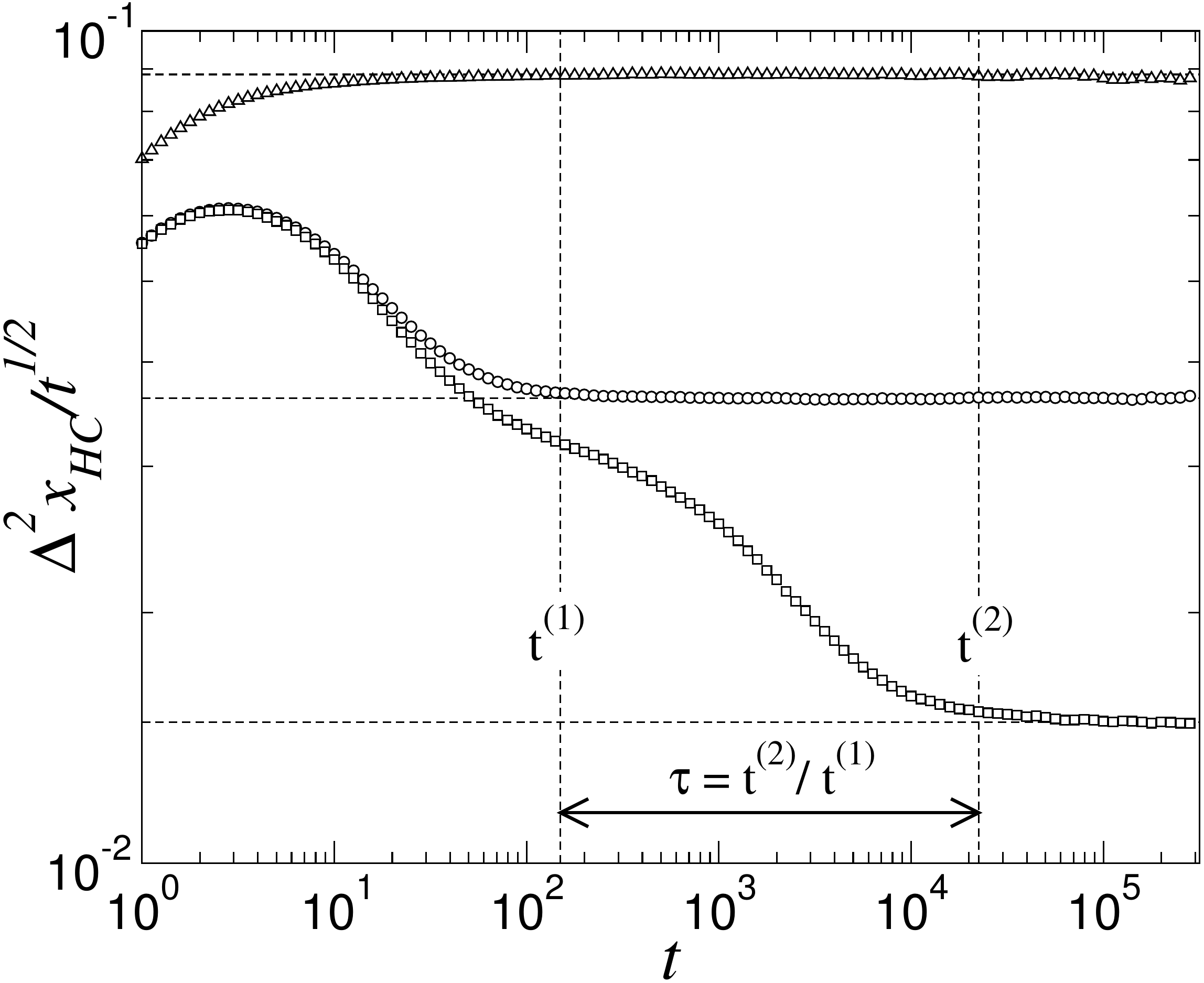}
    \caption{Long time regime of the MSD of a tagged particle in a finite lattice build with zero ($\triangle$), one ($\bigcirc$) and two ($\square$) generations. Parameters: $L=5;\:\lambda=5;\:c=0.9;\:\mbox{and an average over $2\times 10^6$ trajectories.}$ The horizontal dashed lines represents the asymptotic behavior of Eq.~(\ref{msd_hc}) for $n=0,1\:\mbox{and}\:2$ from top to bottom. The vertical dashed lines represent the period $\tau$ given by Eq.~(\ref{tau_cm}).}
    \label{fig:regimen_un_medio}
  \end{center}
\end{figure}

In order to check the validity of Eq.~(\ref{msd_hc}) (where $D^{(n)}_{\mbox{\tiny NI}}$ is given by Eq.~(\ref{coef_cm_hd})) we build the zeroth, first, and second generation of Fig.~\ref{substra} on a finite lattice of $L=5$ sites with periodic boundary conditions. In these cases, Eq.~(\ref{msd_hc}) holds for $t> t^{n}$.
As mentioned above, at very long times, it is expected that the diffusion of a tagged particle reaches a normal behavior (i.e., $\Delta^2 x_{\mbox{{\tiny HC}}} \sim t$) due to  finite-size system effects \cite{beijeren1983}. The simulations were performed up to time $t=3\times 10^5$, and the size of each lattice was chosen large enough in order to avoid the normal regime. In Fig.~\ref{fig:regimen_un_medio}, one can observe a very good agreement between Eq.~(\ref{msd_hc}) and numerical results.

Let us remark that, for the first generation, our results are similar to 
those in Ref.~\cite{Taloni} (single-file diffusion on a periodical potential).
Indeed, if we assume the Arrhenius behaviors  $q_0=A\exp(-E_0/k_BT)$ and $q_1=A\exp(-E_1/k_BT)$,  the MSD of a tagged particle results as in 
Eq.~(\ref{msd_hc}) with $n=1$; and the prefactor of $\sqrt{t}$ takes the form
$\sqrt{D_{\mbox{\tiny NI}}^{(0)}}{\cal G}(d/k_BT)$, where $T$  the absolute temperature, $k_B$ the Boltzmann constant, $E_0$ and $E_1$ energetic barriers, and $d=E_1-E_0$.

\section{Conclusions} \label{sec:conclusion}

We have studied the single-file diffusion problem on one-dimensional self-similar substrates. The system presents a sub-diffusive behavior modulated by log-periodic oscillation. We have found that the global RW exponent of the center of mass $\overline{\nu} (< 1/2)$ is equal to the exponent $\nu$ for a single RW (i.e., when there is only one particle diffusing on the same substrate), while for a tagged particle, the global RW exponent $\nu_{\mbox{{\scriptsize HC}}}$ is equal to $\overline{\nu}/2$. For both quantities, the oscillations occur with the same period $\tau$. The Monte Carlo results are in very good agreement with the values of $\overline{\nu}$ and $\tau$ analytically obtained.

In recent years, many efforts have been dedicated to the study of the single-file diffusion of a tagged particle in a simple one-dimensional lattice where the RW exponent $\nu_{\mbox{{\scriptsize HC}}}$ is $1/4$ \cite{harris,richards1977,beijeren1983,lizana2009,manzi2012,centres2010}. The novelty of this paper is that, by choosing an appropriate value of $\lambda$ for a given $L$, it is possible to design a self-similar structure to obtain a predetermined value of $\nu_{\mbox{{\scriptsize HC}}} (< 1/4)$.
In this case, the resulting diffusive motion will be modulated by logarithmic periodic oscillations with a period $\tau = L^{1/2\nu_{\mbox{{\tiny HC}}}}$.

\section{Aknowledgments}
This work was supported by 
UNMdP and CONICET (PIPs 041 and 431).

\section*{References}

\end{document}